# On Swarm Stability of Linear Time-Invariant Descriptor Compartmental Networks

Cai, Ning[1]    Khan, M. Junaid[2]

[1]College of Electrical Engineering, Northwest University for Nationalities, Lanzhou, China
[2]PN Engineering College, National University of Sciences and Technology, Karachi, Pakistan

**Abstract:** Swarm stability is concerned for descriptor compartmental networks with linear time-invariant protocol. Compartmental network is a specific type of dynamical multi-agent system. Necessary and sufficient conditions for both consensus and critical swarm stability are presented, which require a joint matching between the interactive dynamics of nearest neighboring vertices and the Laplacian spectrum of the overall network topology. Three numerical instances are illustrated to verify the theoretical results.

## 1. Introduction

Compartmental network [1] can be regarded as a specific type of dynamical multi-agent system, which is comprised of special vertices called compartments, interconnected through a network, each containing some substance or information. The neighboring compartments in the network can dynamically exchange the substance or information with each other. Many systems that have been extensively studied in various fields such as biology, chemistry, economics, and engineering can be treated as compartmental systems. For instance, in engineering, some of the artificial neuron networks [2] and sensor networks [3] belong to compartmental networks.

The consensus problem originates from computer science. Its early background is relational to applications such as sensor network data fusion. During the last decade, as a specific type of stability problem, consensus of dynamical multi-agent systems received most extensive attention from the sphere of control theory. Olfati-Saber and Murray [4] pointed that strong connection of the digraph is a sufficient condition for consensus achievement of first-order systems with proper interaction protocol. Ren and Beard [5] proved that the digraph including a spanning tree is the necessary and sufficient condition. Xiao and Wang [6] concerned high-order systems most early and they proposed a criterion based on the structure of certain high dimensional matrices. Wang *et al.* [7] addressed to check whether an appropriate linear high-order consensus protocol exists for a given undirected graph topology. Li *et al.* [8] dealt with the robust stability problem of linear systems with observer type protocols.

[1] Corresponding author: Cai, Ning (E-mail: caining91@tsinghua.org.cn)

The stability of multi-agent systems is different from isolated systems. It has been a common notion that stability of multi-agent systems implicates cohesion. The research on stability problems of multi-agent systems has been conducted for years. Jin *et al.* [9] concerned the stability of a discrete-time system rather early, based on an engineering background of multi-agent supporting system [10]. Later, Liu *et al.* [11] endeavored to extend the study to so-called asynchronous multi-agent systems. Gazi and Passino [12] considered the stability problem of a class of first-order nonlinear models, based on a biological background. Chu *et al.* [13] extended the discussions in [12] to certain anisotropic model. Li [14] also extended the results in [12], concerning the effect of graph topology. Cai *et al.* [15-18] proposed necessary and sufficient conditions for swarm stability of general high order linear [15-17] and nonlinear [18] multi-agent systems, respectively. Recently, Soorki and Tavazoei [19] addressed the asymptotic swarm stability for fractional-order systems.

During recent years, scholars start to notice the stability problems for descriptor multi-agent systems. Descriptor systems are also called singular systems. A descriptor model is more general and precise than a normal model to depict a dynamical physical system, especially as certain algebraic constraints exist among the state variables [20] or as the system dynamics include components with different temporal scales [21]. Xi *et al.* [22-23] early paid attention to the consensus problems of descriptor multi-agent systems, mostly via LMI methods. Yang *et al.* [24] analyzed the consensus conditions for singular multi-agent systems with output feedback protocols. Zhou *et al.* [25] concerned the stability of a class of switching descriptor systems.

This paper is focused on the swarm stability problem of descriptor compartmental networks with LTI dynamical protocols. Necessary and sufficient conditions are offered for both asymptotic swarm stability and swarm stability.

The results in this paper could enrich the stability theory of large-scale descriptor systems. The major contributions are threefold: 1) A criterion to check asymptotic swarm stability is proved via directly studying the structure of the limit of a state transmission matrix, which is a new approach different from those in the literature. 2)

Meanwhile, the proof also provides a typical instance to show the effective application of the almost decouplability conception [17] for directed networks. 3) The swarm stability problem is introduced and discussed in detail, specifically for descriptor compartmental networks. One will see that these results on compartmental networks take much simpler forms than general multi-agent systems.

The organization of the remaining part of this paper is as follows. Section 2 will provide the definition of swarm stability and describe the descriptor compartmental network model. Section 3 will discuss the criteria for checking swarm stability of compartmental networks. Numerical examples will be shown in Section 4. Finally, Section 5 will be the conclusion.

## 2. Problem Description and Preliminaries

### 2.1. Swarm Stability

For dynamical network systems, it is meaningful to redefine the term "stability" since their configuration is essentially different from isolated systems. It has been a common knowledge that stability of a network system implies cohesion, which is formulated by the following definitions.

*Definition 1:* (Swarm Stability) For a dynamical multi-agent system that may be nonlinear and/or time-varying with $x_1, x_2, ..., x_m \in R^n$ the states of $m$ agents, if for $\forall \varepsilon > 0$, $\exists \delta(\varepsilon) > 0$, s.t. $\|x_i(t) - x_j(t)\| < \varepsilon$ ($t > 0$) as $\|x_i(0) - x_j(0)\| < \delta(\varepsilon)$ ($\forall i, j \in \{1, 2, ..., m\}$), then the system is uniformly *swarm stable*. If $\lim_{\varepsilon \to \infty} \delta(\varepsilon) = \infty$, the system is globally uniformly swarm stable.

*Definition 2:* (Asymptotic Swarm Stability) If a dynamical multi-agent system is globally uniformly swarm stable and for $\forall \varepsilon, c > 0$ $\exists T = T(\varepsilon, c) > 0$ s.t. $\|x_i(t) - x_j(t)\| < \varepsilon$ as $t > T(\varepsilon, c)$ and $\|x_i(0) - x_j(0)\| < c$ ($\forall i, j \in \{1, 2, ..., m\}$), then the system is globally uniformly *asymptotically swarm stable*.

Asymptotic swarm stability is also called "consensus" in literature and the two terms will be indiscriminately used hereafter. The above definitions clearly manifest the interior relationship between consensus and stability.

### 2.2. Descriptor compartmental networks

A compartmental network being of *n*th order implies that each vertex may contain

$n$ different types of substance or information, where any type can be transformed into other types. The network's being undirected implicates that if some substance flows from vertex $i$ to $j$, then both the quantities of substance in the two vertices will simultaneously alter. When there exist certain internal algebraic constrains among various quantities, the compartmental network should be depicted by a descriptor model.

A compartmental network can be regarded as a specific type of dynamical multi-agent system. It is supposed to be composed of $m$ vertices indexed from 1 to $m$, each of $m$th order. The state of vertex $i$ is denoted by $x_i = [x_{i1}, x_{i2}, ..., x_{in}]^T \in R^n$, which represents the quantity of substance or information on the vertex. The communication network among vertices is represented by a graph topology $G$ of order $m$. The arc weight of $G$ between vertex $i$ and $j$ is denoted by $w_{ij} \geq 0$, which can be regarded as the strength of communication link. If $w_{ij} > 0$, then vertex $j$ is vertex $i$'s neighbor. The graph can be denoted by its adjacency matrix $W$:

$$G:W = \begin{bmatrix} w_{11} & w_{12} & ... & w_{1m} \\ w_{21} & w_{22} & ... & w_{2m} \\ ... & ... & ... & ... \\ w_{m1} & w_{m2} & ... & w_{mm} \end{bmatrix}$$

The dynamics of the LTI descriptor compartmental network that will be concerned is described by:

$$\begin{cases} E\dot{x}_1 = F\sum_{j=1}^{m} w_{1j}(x_j - x_1) \\ E\dot{x}_2 = F\sum_{j=1}^{m} w_{2j}(x_j - x_2) \\ \vdots \\ E\dot{x}_m = F\sum_{j=1}^{m} w_{mj}(x_j - x_m) \end{cases} \quad (1)$$

where matrices $E, F \in R^{n \times n}$ and $E$ is singular, i.e. $rank(E) < n$. If all state vectors of vertices are stacked together, then the entire state matrix $X \in R^{n \times m}$ of the system is:

$$X = \begin{bmatrix} x_{11} & x_{12} & ... & x_{1m} \\ x_{21} & x_{22} & ... & x_{2m} \\ \vdots & \vdots & & \vdots \\ x_{n1} & x_{n2} & ... & x_{nm} \end{bmatrix}$$

The dynamics of the overall compartmental network (1) can be described by the matrix state equation below:

$$E\dot{X} = -FXL^T \quad (2)$$

where $L = L(G)$ is the Laplacian matrix [14] of the graph topology $G$.

Let $x = \begin{bmatrix} x_1^T & x_2^T & \cdots & x_m^T \end{bmatrix}^T$ be the stack state vector of the system, then the dynamics is depicted by

$$(I_N \otimes E)\dot{x} = -(L \otimes F)x \tag{3}$$

which is the vector form counterpart of (2).

In this model, the matrix $F$ indicates the interactive dynamics between any two neighboring vertices. It intuitively corresponds to the attraction/repulsion relationship in a first order system [4-5]: e.g. if $-F$ is Hurwitz, such a relationship can be regarded as being attractive in an unforced system.

*Lemma 1* (Standard Decomposition of Descriptor Systems): [20] For any regular autonomous descriptor system $E\dot{x} = Ax$ with $E$ being singular, there exist two nonsingular matrices $Q$ and $P$ such that the system is restrictively equivalent to

$$\begin{cases} \dot{x}_1 = A_1 x_1 \\ N\dot{x}_2 = x_2 \end{cases} \tag{4}$$

with the coordinate transformation $\begin{bmatrix} x_1^T & x_2^T \end{bmatrix}^T = P^{-1}x$ ($x_1 \in R^{n_1}$, $x_2 \in R^{n_2}$), $QEP = diag(I_{n_1}, N)$, and $QAP = diag(A_1, I_{n_2})$, where $n_1 + n_2 = n$ and the matrix $N$ is nilpotent.

Usually the first equation in (4) is called a slow subsystem and the second a fast subsystem.

*Definition 3* (Finite Eigenvalue): [20] The set of *finite eigenvalues* for any regular autonomous descriptor system $E\dot{x} = Ax$ or matrix pencil $(E, A)$ is

$$\sigma(E, A)$$
$$= \{\lambda_i(E, A)\}$$
$$= \{s \mid s \in C, |s| < \infty, \det(sE - A) = 0\}$$

It is known that [20] the set of finite eigenvalues $\sigma(E, A)$ equals to the spectrum of matrix $A_1$ in (4), with its cardinal being $|\sigma(E, A)| = n_1$.

*Lemma 2*: [20] The unique solution of a given fast subsystem $N\dot{x}_2 = x_2$ is

$$x_2(t) = -\sum_{i=0}^{h-1} \delta^{(i-1)}(t) N^i x_2(0)$$

where $\delta(t)$ is the delta function and $h$ the nilpotent index of $N$.

# 3. Swarm Stability of Descriptor Compartmental Systems

The major purpose of this section is to prove necessary and sufficient conditions for the swarm stability of LTI descriptor compartmental networks.

The consensus conception is crucial for the study of the dynamics of multi-agent systems. Actually, "Consensus" in dynamical multi-agent systems could be regarded as the counterpart for "equilibrium" in isolated systems [15-16]. Since consensus is essentially a kind of asymptotic stability, it is robust with certain stability margin. So it is of more theoretical importance and will be stressed here, as compared to the critical swarm stability cases.

Before the proof of the criterion to check consensus, i.e. asymptotic swarm stability, several preparations are required to be listed as follow.

*Lemma 3:* The Laplacian matrix $L$ of a directed graph $G$ has exactly a single zero eigenvalue $\lambda_1 = 0$ iff $G$ has a spanning tree, with the corresponding eigenvector $\phi = \begin{bmatrix} 1 & 1 & \cdots & 1 \end{bmatrix}^T$. Meanwhile, all the other eigenvalues $\lambda_2, ..., \lambda_N$ locate in the open right half plane.

*Lemma 4* [16]: If an LTI compartmental network (1) is asymptotically swarm stable, then
$$\lim_{t \to \infty} \dot{x}_i = 0 \quad (i = 1, 2, \cdots, m)$$

*Lemma 5* [26]: For two given matrices $A$ and $B$, let $\lambda(A)$ be an eigenvalue of $A$ with corresponding eigenvector $r(A)$ and $\lambda(A)$ be an eigenvalue of $B$ with eigenvector $r(B)$, then $\lambda(A)\lambda(B)$ is an eigenvalue of $A \otimes B$ with corresponding eigenvector $r(A) \otimes r(B)$.

*Lemma 6* (Almost Decouplability of Directed Network Topology) [17]: For any directed network topology $G$ of $m$ th order and any value $\varepsilon > 0$, there exists a network $G(\varepsilon)$ that is decouplable, i.e. its Laplacian matrix $L(G(\varepsilon)) = [l_{ij}(\varepsilon)]$ is diagonalizable, and meanwhile
$$\sum\nolimits_{i,j=1}^{m} (l_{ij} - l_{ij}(\varepsilon))^2 < \varepsilon$$

*Proposition 1:* The LTI descriptor compartmental network (1) is asymptotically swarm stable if and only if:
1) The network topology $G$ includes a spanning tree;

2) All the values in

$$\{\lambda_i(L)\lambda_j(E,F) | i=2,3,...,m; j=1,2,...,n_1\}$$

have positive real parts, where $\lambda_2(L),...,\lambda_m(L)$ are the nonzero eigenvalues of $L$ and $\lambda_1(E,F),...,\lambda_{n_1}(E,F)$ the finite eigenvalues of $(E,F)$.

*Proof:* Assume that the system is asymptotically swarm stable but $G$ has no spanning tree. Because $G$ has no spanning tree, it must contain $k \geq 2$ different subgraphs $\hat{G}_1, \hat{G}_2,..., \hat{G}_k$, each receiving no information. Let $\xi_1(t)$ and $\xi_2(t)$ denote the consentaneous trajectories of the vertices associated with $\hat{G}_1$ and $\hat{G}_2$, respectively. These two trajectories are independent because they have no information exchange. According to Lemma 2, it can be inferred that $\lim_{t\to\infty}\xi_1(t)$ and $\lim_{t\to\infty}\xi_2(t)$ are constant values. It is possible that

$$\lim_{t\to\infty}\xi_1(t) \neq \lim_{t\to\infty}\xi_2(t) \tag{5}$$

(5) contradicts the assumption that the system is asymptotically swarm stable. Therefore, the condition 1) is necessary.

Now suppose that $G$ includes a spanning tree. In order to solve the differential equation (3), a standard decomposition should first be performed on each vertex according to Lemma 1.

Let $x = (I_m \otimes P)\tilde{x}$, then

$$(I_m \otimes EP)\dot{\tilde{x}} = -(L \otimes FP)\tilde{x}$$

Premultiplying both sides by $I_m \otimes Q$ yields

$$(I_m \otimes QEP)\dot{\tilde{x}} = -(L \otimes QFP)\tilde{x}$$

and

$$(I_m \otimes \begin{bmatrix} I_{n_1} & \\ & N \end{bmatrix})\dot{\tilde{x}} = -(L \otimes \begin{bmatrix} F_1 & \\ & I_{n_2} \end{bmatrix})\tilde{x}$$

In this way, the system (3) is decomposed into the following slow and fast subsystems:

$$\begin{cases} \dot{\tilde{x}}^{(1)} = -(L \otimes F_1)\tilde{x}^{(1)} \\ (I_m \otimes N)\dot{\tilde{x}}^{(2)} = -(L \otimes I_{n_2})\tilde{x}^{(2)} \end{cases} \tag{6}$$

where $\tilde{x}^{(1)} \in R^{mn_1}$ and $\tilde{x}^{(2)} \in R^{mn_2}$.

Let us survey the solution of the fast subsystem above. According to Lemma 6, any directed network topology $G$ could be replaced by a decouplable $G(\varepsilon)$ being arbitrarily close. A sufficiently close $G(\varepsilon)$ should not influence the asymptotic stability [17]. Consider

$$(I_m \otimes N)\dot{\tilde{x}}^{(2)} = -(L(\varepsilon) \otimes I_{n_2})\tilde{x}^{(2)} \tag{7}$$

Suppose that

$$T^{-1}L(\varepsilon)T = diag([0, \lambda_2(L),...,\lambda_m(L)])$$

where $T$ is a nonsingular matrix. Let $\tilde{x}^{(2)} = (T \otimes I_{n_2})\hat{x}^{(2)}$, then (7) converts to

$$(T \otimes N)\dot{\hat{x}}^{(2)} = -(L(\varepsilon)T \otimes I_{n_2})\hat{x}^{(2)}$$

and as a result

$$(I_m \otimes N)\dot{\hat{x}}^{(2)} = -(diag([0, \lambda_2(L),...,\lambda_m(L)]) \otimes I_{n_2})\hat{x}^{(2)} \tag{8}$$

Thus, the fast subsystem (7) is transformed into a series of independent $n_2$ dimensional systems:

$$\begin{cases} N\dot{\hat{x}}_1^{(2)} = 0 \\ N\dot{\hat{x}}_2^{(2)} = \lambda_2(L)\hat{x}_2^{(2)} \\ \quad \vdots \\ N\dot{\hat{x}}_m^{(2)} = \lambda_m(L)\hat{x}_m^{(2)} \end{cases} \tag{9}$$

Because the single zero eigenvalue of $L$ stands for the unobservability of the absolute motion of the network system from the relative measurements [5, 16], the solution of the first equation in (9) is independent of the swarm stability. So it can be left out of our concern. According to Lemmas 2 and 3, it is evident that all the solutions of the rest of equations in (9) tend to be zero as $t \to \infty$.

As to the trajectory of state for the slow subsystem in (6), it can be obtained by solving a normal differential equation, and as a result,

$$\tilde{x}^{(1)}(t) = e^{-(L \otimes F_1)t}\tilde{x}^{(1)}(0) \tag{10}$$

Let us analyze the limit of (10). According to Lemma 5, the $mn_1$ eigenvalues of matrix $-(L \otimes F_1)$ are

$$-\lambda_i(L)\lambda_j(E, F) \quad (i = 1, 2,..., m; \; j = 1, 2,..., n_1)$$

According to Lemma 3, the first $n_1$ values in the spectrum are zero and the remaining ones have negative real parts if the condition 2) holds. Transform $-(L \otimes F_1)$ into Jordan canonical form by nonsingular transformation:

$$-(L \otimes F_1) = UJU^{-1}$$

Evidently, the first $n_1$ eigenvalues of $J$ are zero and the remaining have negative real parts. It follows that

$$\begin{aligned}
&\lim_{t\to\infty} \tilde{x}_1(t) \\
&= \lim_{t\to\infty} e^{-(L\otimes F_1)t}\tilde{x}_1(0) \\
&= U\lim_{t\to\infty} e^{Jt}U^{-1}\tilde{x}_1(0) \\
&= Udiag(\begin{bmatrix}\underbrace{1 \cdots 1}_{n_1} & 0 & 0 & \cdots & \cdots & 0 & 0\end{bmatrix})U^{-1}\tilde{x}_1(0)
\end{aligned} \qquad (11)$$

Let us scan the structure of $\lim_{t\to\infty}\tilde{x}_1(t)$. According to Lemma 5,

$$U = U_L \otimes U_{F_1} = \begin{bmatrix}\alpha_1 & \alpha_2 & \cdots & \alpha_m\end{bmatrix} \otimes \begin{bmatrix}\beta_1 & \beta_2 & \cdots & \beta_{n_1}\end{bmatrix}$$

where $U_L$ and $U_{F_1}$ are nonsingular matrices to transform $L$ and $F_1$ into Jordan canonical forms respectively, with $\alpha_1 \; \alpha_2 \; \cdots \; \alpha_m$ and $\beta_1 \; \beta_2 \; \cdots \; \beta_{n_1}$ their column vectors. Notice that Lemma 5 additionally implies that $\alpha_1 = \begin{bmatrix}1 & 1 & \cdots & 1\end{bmatrix}^T$. So $U$ bears the specific structure

$$U = \begin{bmatrix} \beta_1 & \cdots & \beta_{n_1} & \cdots & \cdots \\ \vdots & & \vdots & & \\ \beta_1 & \cdots & \beta_{n_1} & \cdots & \cdots \end{bmatrix}$$

As a result, (11) leads to

$$\lim_{t\to\infty}\tilde{x}_1(t) = \begin{bmatrix} \beta_1 & \cdots & \beta_{n_1} & 0 & \cdots & 0 \\ \vdots & & \vdots & \vdots & & \vdots \\ \beta_1 & \cdots & \beta_{n_1} & 0 & \cdots & 0 \end{bmatrix}\gamma \qquad (12)$$

where $\gamma = U^{-1}\tilde{x}_1(0)$ denotes an arbitrary vector in $R^{mn_1}$. Evidently, $\lim_{t\to\infty}\tilde{x}_1(t)$ is an $mn_1$ dimensional vector with the values of entries repeating periodically in a cycle of $n_1$ indices. Hence, the system is asymptotically swarm stable.

Contrariwise, the above analysis can be reversed. Assume that system (1) is asymptotically swarm stable. According to Lemma 4, $\lim_{t\to\infty} x(t)$ must be a constant value. Besides, it is an $mn_1$-dimensional vector with the values of entries repeating periodically in a cycle of $n_1$ indices. Therefore, (11) and (12) are true, and it can be seen that the first $n_1$ eigenvalues of $-(L\otimes F_1)$ are zero and the remaining have negative real parts. Thus, all the following values

$$\lambda_i(L)\lambda_j(E,F) \quad (i=1,2,...,m; \; j=1,2,...,n_1)$$

have positive real parts. □

*Remark 1:* Some analogues results to Proposition 1 also exist in the literature, e.g. in [22-24]. The essential highlight here is a novel methodology toward stability analysis of composite systems. The key idea is to analyze the limit of the state transmission matrix and figure out the configuration of system trajectories as $t \to \infty$. Besides, one can sense that these results on compartmental systems are rather concise.

*Remark 2:* Note that it is not necessary for $(E, F)$ to be impulse free here. In this sense, the condition is more relaxed than analogues results in the literature.

According to algebraic theory [26], most of the matrices in $R^{n \times n}$ have all real eigenvalues. Specifically, any real symmetric matrix has spectrum with all real elements. When either $L$ or $F$ possesses all real eigenvalues, the form of Proposition 1 can be dramatically simplified and be formulated as the corollaries below.

*Corollary 1:* If $(E, F)$ has all real finite eigenvalues, then a necessary and sufficient condition for system (1) to be asymptotically swarm stable is that $\lambda_i(E, F) > 0$ $(i = 1, 2, ..., n_1)$ and the network topology $G$ includes a spanning tree.

*Corollary 2:* If the network topology $G$ is symmetric, then a necessary and sufficient condition for system (1) to be asymptotically swarm stable is that $\text{Re}(\lambda_i(E, F)) > 0$ $(i = 1, 2, ..., n_1)$ and $G$ is connected.

The conditions of both corollaries 1 and 2 implicate that the protocol between any neighboring vertices in the network is attractive. This is similar to the swarm stability phenomenon observed by Gazi and Passino [12]. The large-scale system concerned by Gazi and Passino is not asymptotically swarm stable and there is no consensus. The essential reason why their system is free of consensus is that the protocol would become repulsive as two vertices are rather close.

Swarm stability is more general than asymptotic swarm stability, with consensus only a specific type of swarm stability. It is unnecessary for many systems to achieve consensus, but from practical sense, usually a networked system should have swarm stability. There are various such application scenarios in the literature, e.g. flocking [27], formation keeping [28], and containment [29].

The subsequent analysis on swarm stability relies on the next two lemmas.

*Lemma 7* [15-16]: For a normal multi-agent system

$$\dot{x}_i = A x_i + F \sum_{j=1}^{m} w_{ij}(x_j - x_i) \quad (i = 1, 2, ..., m)$$

where $A$ is a matrix with same dimension as $F$ and $\lambda_1 = 0, \lambda_2, ..., \lambda_m \in C$ are the eigenvalues of the Laplacian matrix $L(G)$, if $A$ is stable, then the system is swarm stable iff all the matrices $A - \lambda_i F$ $(i = 1, 2, ..., m)$ are stable, meanwhile, if for some $i$ ($\lambda_i \neq 0$), $A - \lambda_i F$ is critically stable and $L$ is undiagonalizable, each submatrix in $I_m \otimes A - J \otimes F$ corresponding to a Jordan block of $\lambda_i$ which has the form

$$\begin{bmatrix} A - \lambda_i F & -F & 0 & \\ 0 & A - \lambda_i F & \ddots & 0 \\ \vdots & & \ddots & -F \\ 0 & \cdots & 0 & A - \lambda_i F \end{bmatrix}$$

is stable.

*Lemma 8* [21]: The equilibrium point $x = 0$ of $\dot{x} = Ax$ is stable if and only if all eigenvalues of $A$ satisfy $\text{Re}\,\lambda_i \leq 0$ and for each eigenvalue with $\text{Re}\,\lambda_i = 0$ and algebraic multiplicity $q_i \geq 2$,

$$rank(A - \lambda_i I) = n - q_i$$

where $n$ is the dimension of $x$.

*Proposition 2:* For the descriptor compartmental network (1) with $\{\lambda_1 = 0, \lambda_2, ..., \lambda_m\}$ being the spectrum of $L(G)$, suppose that $(E, F)$ is regular and pulse free, then the overall system is swarm stable iff all the matrix pencils $(E, -\lambda_i F)$ ($i = 1, 2, ..., m$) achieves admissible stability, meanwhile, $(E, F)$ has no finite eigenvalues on the imaginary axis except zero if $L$ is undiagonalizable.

*Proof:* According to the analysis in the proof of Proposition 1, the admissible swarm stability is determined by the slow subsystem in (6). According to Lemma 7, the slow subsystem is swarm stable iff all the matrices $-\lambda_i F_1$ are stable, meanwhile, each Jordan block of $-J_L \otimes F_1$ with the form

$$\begin{bmatrix} -\lambda_i F_1 & -F_1 & 0 & \\ 0 & -\lambda_i F_1 & \ddots & 0 \\ \vdots & & \ddots & -F_1 \\ 0 & \cdots & 0 & -\lambda_i F_1 \end{bmatrix} \quad (13)$$

is also stable if $-\lambda_i F_1$ is critically stable, with $J_L$ the Jordan canonical form of $L$.

Suppose that $J_{F_1} = H^{-1} F_1 H$, where $J_{F_1}$ is the Jordan canonical form of $F_1$ and $H$ the corresponding nonsingular matrix. The stability of (13) is equal to that of

$$(I \otimes H^{-1}) \begin{bmatrix} -\lambda_i F_1 & -F_1 & 0 & \\ 0 & -\lambda_i F_1 & \ddots & 0 \\ \vdots & & \ddots & -F_1 \\ 0 & \cdots & 0 & -\lambda_i F_1 \end{bmatrix} (I \otimes H)$$

$$= \begin{bmatrix} -\lambda_i J_{F_1} & -J_{F_1} & 0 & \\ 0 & -\lambda_i J_{F_1} & \ddots & 0 \\ \vdots & & \ddots & -J_{F_1} \\ 0 & \cdots & 0 & -\lambda_i J_{F_1} \end{bmatrix} \quad (14)$$

According to Lemma 8, (14) is stable iff $-\lambda_i J_{F_1}$ is stable and

$$rank([-\lambda_i J_{F_1} - \lambda I_{n_1}, \ -J_{F_1}]) = rank(-\lambda_i J_{F_1} - \lambda I_{n_1}) \quad (15)$$

where $\lambda$ is any eigenvalue of matrix $-\lambda_i J_{F_1}$ with zero real part. It is obvious that if $\lambda \neq 0$, (15) cannot hold. Assume that $(E, F)$ has some nonzero finite eigenvalues on the imaginary axis, which is denoted by $\pm \gamma j$ ($\gamma > 0$), and let $\lambda_i = \alpha + \beta j$. If $\beta = 0$, then $\lambda \neq 0$; whereas if $\beta \neq 0$, $(E, -\lambda_i F)$ must possess a finite eigenvalue with positive real part. So $(E, F)$ should not have any nonzero finite eigenvalues on the imaginary axis. According to the relevant theory about linear descriptor systems [20], the statement of the current proposition is true as a necessary and sufficient condition for swarm stability. □

*Remark 3:* The criteria provided by Propositions 1 & 2 possess evident theoretical advantages. First, they convert the stability problem of a large-scale descriptor system into the stability of a series of low-dimensional matrix pencils and significantly reduce the computational complexity. Second, if a system is to be stabilized, the requirements for the topology of network is separated from the requirements for the interactive dynamics between vertices.

*Remark 4:* The form of criteria to check swarm stability for system (1) is very simple, as comparing with the criteria for other types of similar models [22-25].

So far as the swarm stability is concerned for compartmental network, there is no requirement for the connectivity of the network. This differs from the typical notion in the literature about consensus that the network topology usually should include a spanning tree. The reason is that the vertices have no dynamics by themselves, with their motions in the state space only determined by the interactions. If any vertex is isolated, without any neighbor, then its state will never alter.

The criteria provided by Propositions 1 & 2 for checking swarm stability will be visually illustrated by the examples in the next section.

## 4. Simulations

In this section, numerical instances will be exhibited to illustrate the theoretical results. The network topologies that will be concerned are shown in Fig.1 with

$$W = \begin{bmatrix} 0 & 0 & 1 & 0 & 0 \\ 1 & 0 & 0 & 0 & 0 \\ 1 & 1 & 0 & 0 & 1 \\ 1 & 1 & 1 & 0 & 0 \\ 1 & 0 & 0 & 0 & 0 \end{bmatrix}$$

in (a) and

$$W = \begin{bmatrix} 0 & 2 & 0 & 2 & 1 \\ 2 & 0 & 0 & 0 & 0 \\ 2 & 2 & 0 & 0 & 1 \\ 0 & 2 & 1 & 0 & 1 \\ 0 & 0 & 0 & 1 & 0 \end{bmatrix}$$

in (b). Both of them have spanning trees.

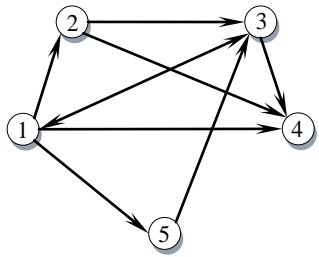

(a)

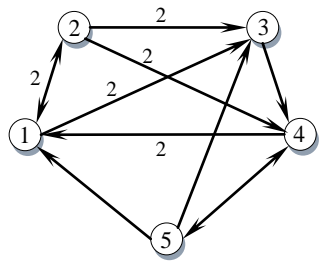

(b)

Fig.1. Network topologies. 1 is default weight of arc.

Suppose that

$$E = \begin{bmatrix} 2 & 2 & 0 \\ -1 & -2 & 1 \\ 1 & 3 & -2 \end{bmatrix}$$

throughout the current subsection. Its rank is 2, so all instances are singular.

In the first instance, let the network topology be the one shown in Fig. 1, with

$$F = \begin{bmatrix} 2 & 2 & 2 \\ 0 & 0 & 1 \\ 0 & 1 & 2 \end{bmatrix}$$

The finite eigenvalues are 1 and 0.1667, meanwhile, the spectrum of the Laplacian matrix is $\{0, 1.2679, 5.5 \pm 1.3229i, 4.7321\}$. According to Proposition 1, the system is asymptotically swarm stable. The simulation result is shown in Fig. 3, with $t \in [0,12]$ and

$$X(0) = \begin{bmatrix} -3.2776 & -5.9099 & 1.8076 & 1.5806 & -2.9512 \\ -0.4794 & 3.6798 & -1.8174 & -3.8132 & 0.9020 \\ -0.4735 & -1.6135 & -0.9556 & -0.1104 & -1.0944 \end{bmatrix}$$

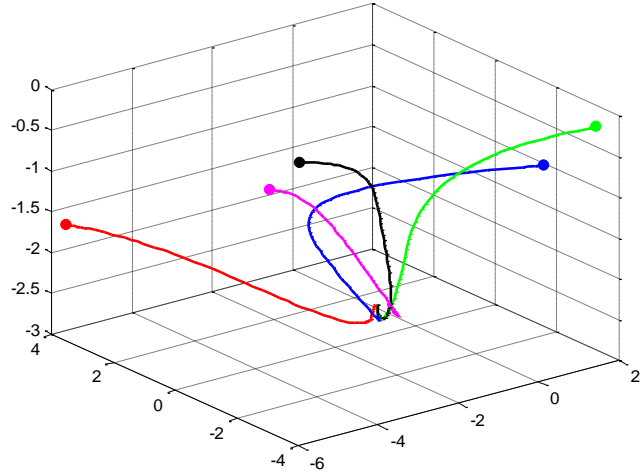

Fig.2. Consentaneous trajectories of first instance. ($t \in [0,12]$)

Thick dots are starting points.

In the second instance, let the network topology be the one shown in Fig. 1, with

$$F = \begin{bmatrix} 1 & 2 & 0 \\ 1.125 & 1 & 0 \\ 0 & 0 & 10 \end{bmatrix}$$

The finite eigenvalues are $\pm 0.7809i$. The Laplacian matrix is

$$L = \begin{bmatrix} 1 & 0 & -1 & 0 & 0 \\ -1 & 1 & 0 & 0 & 0 \\ -1 & -1 & 3 & 0 & -1 \\ -1 & -1 & -1 & 3 & 0 \\ -1 & 0 & 0 & 0 & 1 \end{bmatrix}$$

with spectrum $\{0,1,2,3,3\}$, which is undiagonalizable. The system must be swarm unstable according to Proposition 2. The simulation result is shown in Fig. 3, with $t \in [0,6]$ and

$$X(0) = \begin{bmatrix} -5.9138 & -1.1723 & -2.0973 & -0.9124 & -1.7271 \\ 2.5077 & 0.7047 & 1.7981 & -0.0531 & 3.9337 \\ 3.2370 & 2.4445 & 2.3951 & 2.6190 & 1.6665 \end{bmatrix}$$

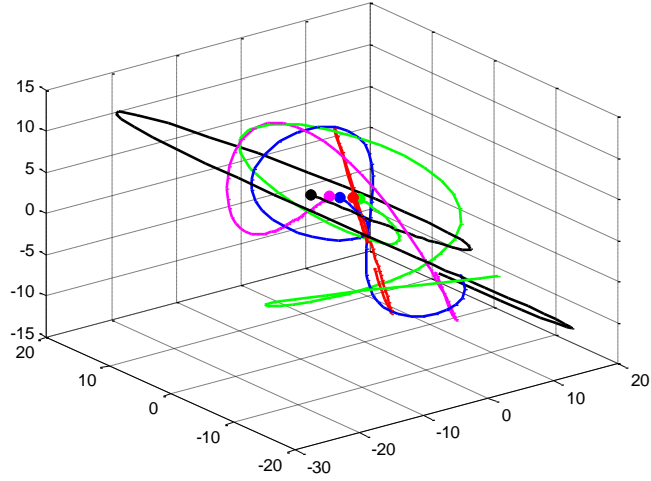

Fig. 3. Swarm unstable trajectories of second instance. ($t \in [0,6]$)

In the third instance, keep $E$ & $F$ and just let the weighted network topology have a slight variation as

$$W = \begin{bmatrix} 0 & 0 & 1 & 0 & 0 \\ 1 & 0 & 0 & 0 & 0 \\ 1 & 1 & 0 & 0 & 1 \\ 1 & 1 & 2 & 0 & 0 \\ 1 & 0 & 0 & 0 & 0 \end{bmatrix}$$

This time the Laplacian matrix is diagonalizable with the spectrum $\{0,1,2,3,4\}$. The system is now swarm stable according to Proposition 2. The simulation result is shown in Fig. 4, with $t \in [0,8]$ and

$$X(0) = \begin{bmatrix} 4.8626 & 8.1001 & -0.3978 & -3.2563 & 1.7169 \\ 2.6541 & -7.5692 & -4.3644 & 1.9946 & -0.7866 \\ -1.6885 & 0.4988 & 1.8697 & 0.7470 & 0.2093 \end{bmatrix}$$

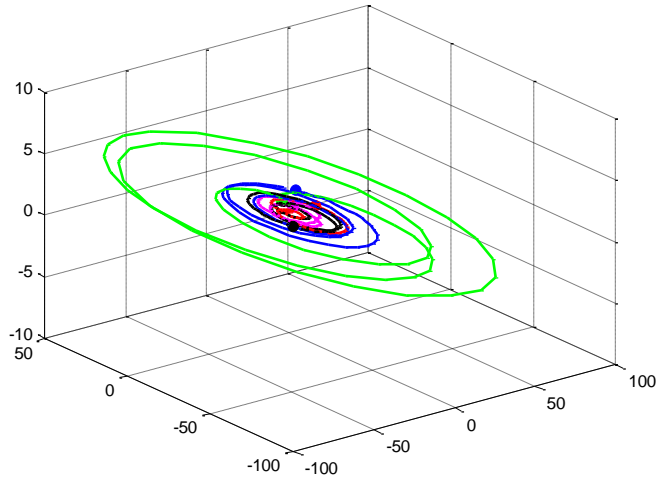

Fig. 3. Critical swarm stable trajectories of third instance. ($t \in [0,8]$)

*Remark 5:* It is interesting that if the weighted network topology bears uncertainty, then the probability for the system above to be swarm stable is nearly equal to 1, because undiagonalizable Laplacian matrix is actually a rather special case among any arbitrary weighted networks [17]. Such a fact might seem weird to contradict the common sense that critical stability is not robust.

## 5. Conclusion

This paper deals with the swarm stability problem of descriptor compartmental networks with LTI dynamical protocol. The background of this problem is from various application fields. The conception of swarm stability is formally defined concerning cohesion. Necessary and sufficient conditions for the swarm stability of descriptor compartmental networks are proved, based on studying the structure of the analytical solution of high-order state equation. The conditions require a joint matching between the finite eigenvalues of single subsystem and the Laplacian spectrum of the overall network topology. One may see that these criteria are quite simple. Numerical instances are shown to illustrate the theoretical results. In the future, the current research might be further extended under variant directions, such as considering systems with certain nonlinearity via the technique of transforming the models into canonical forms; or conducting a deeper exploration to the robustness of swarm stability.

# Acknowledgements

This work is supported by National Natural Science Foundation (NNSF) of China (grants 61174067, 61263002, & 61374054).